\documentclass[journal]{IEEEtran}
\usepackage{cite}

\ifCLASSINFOpdf
\else
\fi
\usepackage{tabularx,booktabs}
\usepackage{multicol}
\usepackage{caption}
\usepackage{subcaption}
\usepackage{cuted}
\usepackage{xpatch}
\usepackage{graphicx}
\usepackage[caption=false,font=footnotesize]{subfig}
\usepackage{amssymb}
\usepackage{enumitem}
\usepackage{epstopdf}
\usepackage{float}
\usepackage{algorithmic}
\usepackage{array}
\usepackage{amsmath}
\usepackage{amssymb}
\usepackage{amsthm}
\usepackage{diagbox}
\usepackage{mdwmath}
\usepackage[table]{xcolor}
\usepackage{makecell}
\usepackage{eqparbox}
\usepackage{stfloats}
\usepackage{fixltx2e}
\usepackage{tabularx}
\usepackage{cases} 
\usepackage{booktabs}
\usepackage{flushend}
\usepackage{threeparttable}
\usepackage{xcolor}
\usepackage{makecell}
\usepackage{multirow}
\definecolor{headergray}{gray}{0.85}
\definecolor{lightgreen}{RGB}{220,245,220}

\usepackage[boxed,ruled,commentsnumbered]{algorithm2e}
\usepackage{url}
\usepackage{cite}
\allowdisplaybreaks[4]

\makeatletter

\newcommand{\Rmnum}[1]{\expandafter\@slowromancap\romannumeral #1@}
\makeatother

\floatname{algorithm}{Procedure}

\graphicspath{{./picture/} }
\usepackage{caption}
\usepackage{booktabs}   
\usepackage{array}      
\usepackage{multirow}   
\usepackage{booktabs}       
\usepackage{tabularx}       
\usepackage{siunitx}        
\usepackage{multirow}       
\usepackage{geometry}       
\begin{document}
%
\title{Stratospheric Grid: A Wireless Power Transfer Enabled HAP Network with Integrated Generation–Grid–Load–Storage Functions}
\author{Peng Wang, Eros Kuikel, Jia Ye, Mohamed-Slim Alouini%
}
%

\markboth{Journal of \LaTeX\ Class Files,~Vol.~, No.~, ~}%
{Shell \MakeLowercase{\textit{et al.}}: Bare Demo of IEEEtran.cls for IEEE Journals}
%



\maketitle

\begin{abstract}
Conventional high-altitude platforms (HAPs) face challenges in achieving continuous all-weather operation due to intermittent photovoltaic power generation, limited energy storage capacity, and high mission loads resulting from functional integration. To address this fundamental issue, we propose a stratospheric energy grid in which wireless power transfer (WPT) interconnections constitute the grid layer, while HAPs operate as dynamically reconfigurable integrated generation–grid–load–storage (IGGLS) nodes that harvest, buffer, consume, and peer-to-peer transfer energy for constellation-level balancing and resilience. In this system, each HAP node can flexibly switch among energy source, load, and storage roles according to its energy status and mission requirements, enabling energy exchange and spatiotemporal optimization within the stratosphere. Through cooperative scheduling, the stratospheric grid not only enables surplus-to-deficit energy support among HAPs but also extends upward to satellites and downward to the terrestrial grid and communication infrastructure, forming a heterogeneous, WPT-mediated interconnection. As an IGGLS ecosystem, it exploits peer-to-peer energy logistics, spatiotemporal smoothing of intermittency, cross-domain backup via the terrestrial grid, and service-aware dispatch, thereby boosting overall energy utilization and operational resilience. The proposed approach is validated through case studies, and we delineate an agenda of feasible research directions.
\end{abstract}

\IEEEpeerreviewmaketitle

\section{Introduction}
The concept of high altitude platforms (HAPs) was first proposed in the 1990s, originating from the idea of deploying stratospheric aerial vehicles such as balloons, airships, or solar-powered unmanned aircraft to establish quasi-satellite communication relay systems in areas where terrestrial infrastructure was difficult or costly to deploy \cite{tozer2001high}. The primary goal was to achieve wide-area coverage, low latency, and rapid deployment of communication and monitoring services at altitudes ranging from 20 to 50 km, effectively bridging the gap between terrestrial networks and satellites. With advances in flight control, lightweight energy systems, and high-efficiency communication payloads, the role of HAPs has evolved from a simple communication relay to a multi-functional platform integrating communication, sensing, navigation, and computing. In today’s space–air–ground-integrated network (SAGIN) framework, HAPs are recognized as crucial mid-layer nodes that connect space and ground networks and can establish optical or microwave links with satellites while providing adaptive coverage and edge computing to ground and aerial terminals. Consequently, HAPs are transforming from traditional auxiliary communication facilities into intelligent, reconfigurable aerial base stations, playing a pivotal role in future 6G systems and emerging low-altitude economy applications.

However, HAP operations are fundamentally constrained by onboard energy supply. Early platforms relied on conventional sources such as fuel tanks and batteries, which limited endurance to roughly 48 hours and necessitated frequent landings for refueling \cite{karapantazis2005broadband}. With advances in lightweight structures, high-efficiency photovoltaics, and power management, most modern HAP concepts adopt solar energy as the main source. Operating above the clouds provides abundant sunlight, and the large surface area of HAP airframes accommodates extensive solar arrays that generate substantial power. To ensure continuity through night periods and winter seasons, solar-powered HAPs pair their arrays with secondary storage or carriers such as rechargeable batteries or hydrogen fuel cells, which are replenished during daytime flight. Despite substantial advances in HAP energy systems, the operational energy landscape still exhibits two structural mismatches.

\textbf{Contradiction 1: The underutilization of daytime solar generation.} In the stratosphere, abundant and long-duration sunlight allows the onboard solar panels to reach peak output around midday. However, the rhythm of energy generation does not align with the rhythm of network demand. During these hours, most operations are below the maximum output of photovoltaics, leading to a mismatch between generation and consumption, and consequently, curtailment or wastage of renewable energy. Moreover, when the onboard batteries approach full charge, the excess power cannot be effectively stored, further amplifying the redundancy.

\textbf{Contradiction 2: The nighttime or high-load energy shortage.} During hours without sunlight or in short-term high-demand periods such as heavy backhaul traffic, dense beamforming, or edge computing, the stored energy alone is often insufficient to sustain long-duration and high-power operation. In addition, long nights and low temperatures at high latitudes reduce available battery capacity, while strong winds increase propulsion power consumption. These factors compound during peak service periods, creating a “power gap” that can result in link downgrades, degraded service quality, or even complete outage.

In essence, temporal and magnitude mismatches between generation and demand depress renewable energy utilization and raise energy shortage risks during critical service windows. Addressing this dual constraint requires a system level approach that ensures sustainable, resilient, and highly available power delivery and, by extension, reliable HAP communication services. Building on these observations, the guiding idea is to mobilize energy rather than strand it, routing daytime surpluses from platforms with excess generation to nodes facing immediate or anticipated deficits and providing timely top ups during nocturnal or peak load intervals. To realize this across a dispersed stratospheric fleet, a practical enabler is wireless power transfer (WPT), which can coordinate point-to-point energy flows through spatial fields without relying on traditional physical conductors. 

In fact, applying WPT to HAPs is not a new idea. As early as the 1980s, researchers proposed microwave power beaming, exemplified by the SHARP program, which used a large ground antenna to transmit wide-aperture microwave beams to airborne receivers equipped with rectifying antenna arrays that converted RF energy into DC power \cite{brown1986microwaver}. Parallel efforts explored laser power beaming, with proof-of-concept experiments in Japan and the United States demonstrating the possibilities of laser-powered HAP flight. 
Although these studies established the experimental feasibility of energy transfer to and from HAP platforms, what remains missing is an energy-centric view that positions HAPs as cooperative nodes within a stratospheric network. DARPA’s Persistent Optical Wireless Energy Relay (POWER) program explicitly frames this vision as a resilient “wireless energy web,” where speed-of-light power routing over multi-path links provides rapid reconstitution and graceful degradation under disruption \cite{POWER_SPIE}. Its objective architecture builds high-altitude relay nodes that \emph{redirect}, \emph{correct} (via adaptive optics), and \emph{selectively harvest} energy before beaming it to loads for conversion, effectively treating generation–distribution–consumption as an interconnected web of transactions \cite{POWER_SPIE}.

In 2021 the terrestrial power grid systematically introduced the novel integrated generation-grid-load-storage (IGGLS) architecture to address system stability and renewable energy integration challenges caused by high penetration of renewable energy\cite{10158921}. Inspired by the IGGLS paradigm, HAPs can utilize WPT to form a stratospheric grid. This is enabled by their inherent design as integrated platforms, which combine communication, sensing, and propulsion systems (load), photovoltaic power generation (generation), and onboard batteries (storage). Within this architecture, platforms reconfigure roles dynamically, while predictive dispatch co-optimizes energy routing with coverage and traffic objectives, thereby anticipating deficits, flattening peaks, and mitigating daytime renewable curtailment. In this article, we reconceptualize the HAP network from the IGGLS perspective and propose feasible implementation strategies. We further analyze its distinctive characteristics, operational potential, and application scenarios, thereby formalizing the stratospheric energy grid as a system-level architecture. Furthermore, we outline the prospective application scanerios toward realizing an integrated, intelligent, and sustainable energy–communication framework that connects space, air, and ground systems to support continuous and resilient 6G-era connectivity.

\begin{table*}[htbp]
\centering
\caption{Comparison of Source-Grid-Load-Storage Configurations}
\label{table:comparison}
\begin{tabular}{p{1.3cm} p{4 cm} p{4 cm} p{6cm}}
\toprule
\textbf{Component} & \textbf{Stratospheric Grid} & \textbf{Conventional Grid} & \textbf{Key Differences} \\
\midrule
\textbf{Generation} &
\begin{itemize}[leftmargin=*,nosep]
    \item Mobile solar generation nodes
    \item Distributed generation units
    \item Platform-mounted PV systems
\end{itemize} &
\begin{itemize}[leftmargin=*,nosep]
    \item Fixed power plants
    \item Centralized generation
    \item Various types of power plants
\end{itemize} &
Generation units shift from \textbf{fixed and centralized} to \textbf{mobile and distributed}, with power generation capability strongly dependent on platform position and attitude. \\
\rowcolor[gray]{0.95}
\textbf{Grid} &
\begin{itemize}[leftmargin=*,nosep]
    \item WPT interconnection links
    \item Mobile energy hubs
    \item Predictive, distributed energy routing layer 
\end{itemize} &
\begin{itemize}[leftmargin=*,nosep]
    \item Wired transmission lines
    \item Fixed substations
    \item Predefined power routing and hierarchical control
\end{itemize} &
The energy network evolves from a \textbf{wired and fixed} physical infrastructure to a \textbf{wireless and dynamically formed} virtual architecture. \\
\textbf{Load} &
\begin{itemize}[leftmargin=*,nosep]
    \item Mission payloads (communication, sensing, computing)
    \item Housekeeping/avionics and thermal management
    \item Propulsion and actuation power
\end{itemize} &
\begin{itemize}[leftmargin=*,nosep]
    \item Residential demand
    \item Commercial and industrial demand
    \item Emerging flexible load (Electric Vehicles, data centers)
\end{itemize} &
Load characteristics shift from \textbf{user-behavior-driven} consumption to \textbf{platform-mission-driven} functional requirements. \\
\rowcolor[gray]{0.95}
\textbf{Storage} &
\begin{itemize}[leftmargin=*,nosep]
    \item Onboard storage modules
    \item WPT-mediated virtual storage pooling
    \item Mission/peak-power buffering
\end{itemize} &
\begin{itemize}[leftmargin=*,nosep]
    \item Utility-scale storage
    \item Substation/behind-the-meter storage
    \item Long-duration storage and market services
\end{itemize} &
Storage evolves from \textbf{centralized, stationary grid resources} to \textbf{distributed, mobile platform units, aggregated virtually through WPT} for mission-oriented buffering and constellation-level balancing. \\
\bottomrule
\end{tabular}
\end{table*}    


\section{IGGLS based Stratospheric Grid}
In the traditional IGGLS framework, generation denotes the production side of the power system, meaning generation resources such as thermal power, hydropower, nuclear power, and increasingly renewable sources like solar and wind. Grid refers to the transmission and distribution network that serves as the backbone connecting generation to end users, including high-voltage transmission lines, regional substations, and local distribution systems. Load represents the system’s electrical demand, encompassing residential, industrial, and commercial consumers as well as emerging sectors such as electric vehicles and data centers. Storage comprises energy storage systems that temporarily hold and later release energy to balance fluctuations between supply and demand, including batteries, pumped hydro, and compressed-air systems. Viewed against this framework, a HAP inherently exhibits generation, load, and storage functions, where solar arrays act as power resources, communication and propulsion subsystems constitute loads, and onboard batteries or hydrogen fuel cells provide storage. WPT constitutes the grid layer by interlinking platforms and enabling coordinated power flows across the constellation. In this way, the conceptual foundation can be translated to the stratospheric domain, forming an aerial power grid. Table \ref{table:comparison} summarizes the composition and characteristics of conventional terrestrial and stratospheric power grids across four dimensions, and highlights the similarities and differences in their respective architectures. On this basis, the following text will delve into the physical foundations that make up the IGGLS based stratospheric grid.

\subsection{Generation in Stratospheric Grid}
Operating at 17--25 km in the stratosphere confers distinct advantages for solar power. Above Stratospheric weather, scattering and absorption are greatly reduced, the shorter atmospheric path yields higher irradiance than on the ground, and the platform can enjoy extended daylight windows that can be further lengthened through mild maneuvering or deployment in high-insolation regions. Using the top-of-atmosphere solar constant of about $1361~\mathrm{W/m^2}$ as a reference, HAPs can realize peak direct normal irradiance around $1100~\mathrm{W/m^2}$ near the equator in winter or with modest zenith offsets, and up to about $1250~\mathrm{W/m^2}$ under favorable geography and attitude when the array is normal to the Sun.

Because air density is low and lift margins are tight, HAPs must be extremely lightweight. Large lifting surfaces therefore double as the primary locations for photovoltaic arrays, manifesting in the form of high-aspect-ratio wings on heavier-than-air platforms and the outer envelope on lighter-than-air vehicles. Under these constraints, specific power takes precedence over absolute conversion efficiency. Therefore, onboard photovoltaic systems widely employ flexible multi-junction III-V cells (with laboratory efficiencies exceeding 40\%) and lightweight thin-film technologies (such as CIGS and organic cells) to optimize the power-to-mass ratio. Although the efficiency of organic cells ($\approx 8.7\%$) currently lags behind that of silicon-based ($>20\%$) or GaAs ($>30\%$) cells, their ultra-lightweight and conformable characteristics enable the maximization of the effective power-generating area. This ultimately transforms the aircraft fuselage into a mobile, aerial power station.

For practical quantification, assuming an end-to-end system efficiency of approximately 20\%, peak electrical output is primarily determined by the area of the photovoltaic array. As shown by the representative platforms in Table \ref{tab:haps_power_analysis}, a small HAP generates roughly $1.1-2.5$ kW, which is sufficient to support low-thrust propulsion and light payloads. Medium and large HAPs produce around $6.6-12.5$ kW and $17.6-22.5$ kW, respectively, enabling them to support more power-demanding missions, such as 5G communications. Recently, the HAPSMobile HAWK30, developed by Softband and AeroVironment, has emerged as one of the largest platforms, featuring a photovoltaic array of about $219 m^2$ and delivering peak power of approximately $48-55$ kW. This capability enables multi-kilowatt continuous energy supply and rapid daytime charging, ensuring sufficient power for long-night operations. The HAWK30 platform clearly demonstrates the potential to evolve into a significant aerial power hub for future missions. 
\begin{figure*}[ht] 
\centering
\includegraphics[width=0.9\textwidth]{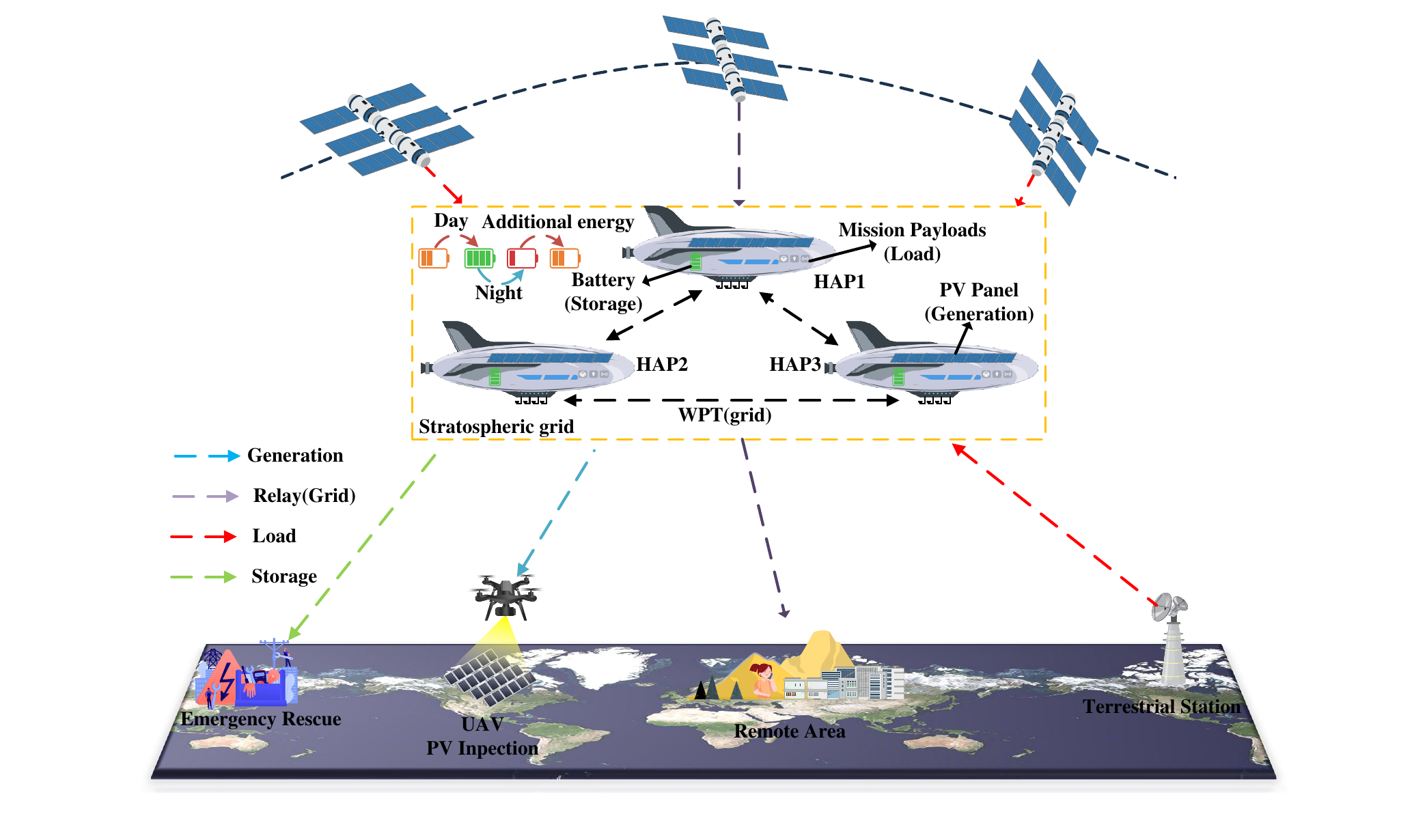} 
\caption{Architecture of the stratospheric grid.}
\centering
\label{fig:1}
\end{figure*}

\subsection{Grid in Stratospheric Grid}
WPT is regarded as a key pathway to overcoming the energy bottleneck of HAPs. Within the IGGLS framework, it serves as the “grid”, enabling on-demand energy replenishment, relay-based power sharing, and centralized energy delivery among HAPs in the stratosphere. This capability extends mission duration and supports higher-power payloads. From a physical standpoint, WPT can be categorized into two main types, namely far-field transmission based on electromagnetic radiation and near-field transmission based on non-radiative electromagnetic coupling. Since near-field methods are only effective over centimeter-scale distances, they are not feasible for stratospheric power networks. Therefore, the following discussion focuses on radio-frequency and laser-based far-field transmission mechanisms and how they enable the “grid” functionality within the stratosphere.

\subsubsection{RF-WPT}
In the case of RF-WPT, thin high-altitude cloud and aerosol loading contributes negligible excess loss, ensuring resilient operation. In the millimeter-wave bands, the short wavelength yields higher gain and narrower beamwidth for a given antenna aperture. Very high transmit gain can offset large free-space path loss over long ranges, making mmWave a prime candidate for stratospheric energy transfer between HAPs. Atmospheric absorption by oxygen and water vapor and rain attenuation in the stratospheric still matter, so operation should favor spectral windows outside major absorption lines, e.g., 5.8 GHz, 28–48 GHz, and 94 GHz.

Both ends typically employ large digital or hybrid phased arrays to form ultra-narrow beams and maintain precise pointing. End-to-end efficiency is strongly shaped by the receive-side RF-to-DC stage. At \({95~\mathrm{GHz}}\), GaN/SiC voltage-doubler rectenna circuits have achieved up to \({61.5\%}\) RF-to-DC efficiency and demonstrated tolerance to high power density (about \({3.6~\mathrm{kW/m^2}}\)). To meet the mass and volume constraints of HAPs, rectifying antenna arrays favor integrated designs such as compact circularly polarized substrate integrated waveguide cavity-backed arrays with self-biased rectifiers. Circular polarization helps mitigate fluctuations in received power caused by small attitude errors.

On the other hand, HAP separations are typically tens to hundreds of kilometers, so RF-WPT is sensitive to free-space path loss. However, with meter- to tens-of-meters-scale apertures and mmWave operation, the radiating near-field (Fresnel) region can extend to tens of kilometers. For instance, using the Fresnel distance rule, an aperture of 10 to 15 meters at a wavelength of 5 to 10 millimeters gives a Fresnel region of about 20 to 90 kilometers. In this region, the field is still radiative but not well approximated by a plane wave. Geometric loss is lower than that in the far field with received power falling with the square of distance, which can raise transfer efficiency and is useful for formation flight energy relay or concentrated supply. 

\subsubsection{Optical WPT}
In the context of optical WPT, high-power laser beams transport energy from source to receiver. Because the optical wavelength is much shorter than that of RF waves, it can theoretically achieve higher energy concentration and narrower beam divergence, thereby maintaining a high power density over longer distances. At stratospheric altitudes, the air is thin and contains little water vapor or aerosol, resulting in very low transmission loss. However, when the link extends across regions with strong atmospheric density gradients, turbulence can induce random fluctuations, beam spreading, and scintillation. The resulting angular jitter can reach or even exceed the microradian scale, surpassing the angular tolerance of the receiver’s aperture and causing severe power loss or link interruption. Therefore, the pointing, acquisition, and tracking (PAT) subsystem of optical WPT must incorporate adaptive optics to dynamically correct wavefront distortion and suppress jitter in real time. For instance, DARPA’s POWER program aim to use wavefront sensing and adaptive-optics correction within the relay, since their demonstration involves long-range optical links over small apertures on vibrating airborne platforms\cite{POWER_SPIE}.

The system architecture typically consists of a high-efficiency laser array as the transmitter, a large-aperture telescope for beam collimation and fine focus adjustment, and a dedicated photovoltaic array on the receiver side for energy conversion. Common operating wavelengths are in the near-infrared range (e.g., 800 nm or 1550 nm), which offer favorable atmospheric transmission windows and relatively safe eye exposure levels. The receiver can employ narrow bandgap photovoltaic converters such as indium gallium arsenide that are spectrally matched to the laser wavelength to maximize photoelectric conversion efficiency. Alternatively, conventional solar cells may be reused to simplify engineering implementation, though at the cost of lower conversion efficiency and the need for additional spectral management.

Several European startups have demonstrated shorter-range optical WPT links that closely parallel HAP-scale beam geometries. SunCubes achieved laser charging of UAVs over 600\,m and is scaling toward 3\,km with auto-tracking diode arrays and lightweight photovoltaic receivers, while Powerin.Space completed 1.5\,km stabilized laser links backed by ESA BIC funding. Both projects validate continuous airborne laser refueling and low-jitter beam tracking under motion~\cite{suncubes-web,powerinspace1.5km}. Recent field tests have demonstrated optical power beaming at kilowatt-class levels over multi-kilometer paths. For example, 800 W was delivered across 8.6 km using a near-infrared fiber laser, achieving about 20\% end-to-end efficiency with a compact tiled PV receiver \cite{nwpu_dronenews,fivemilenews,darpa_newatlas}. In terms of air to ground energy transfer experiments, Japan Space Systems steered a 5.8~GHz power beam from a fast aircraft to a ground receiver over 5~km using a phased-array transmitter \cite{japan-success}.

In fact, the two modalities are complementary within a stratospheric power network. Optical links are preferred under clear-sky, long-range, and well-aligned conditions, while RF links are more suitable when traversing thin clouds, encountering light aerosols, or requiring wider pointing tolerances. In formation flight and cross-layer energy scheduling, adaptive switching and joint coordination between the two link types can be achieved through real-time awareness of weather and geometric states, ensuring continuous and stable network operation.  


\subsection{Load in Stratospheric Grid}
The “load” generally comprises the power for station-keeping, and the power required by mission payloads.

\textbf{Station-keeping:} Propulsion is the dominant component of the auxiliary energy budget, and the required power grows approximately with the cube of airspeed. The operating point is further shaped by platform mass, by aerodynamic efficiency as captured by the lift-to-drag ratio, and by the efficiency of the propulsion chain. In calm conditions, a small platform can sustain cruise on only a few hundred watts, yet the demand rises rapidly in typical mid-latitude winds as the cubic dependence asserts itself. For instance, a Aquila-class vehicle with a mass of about 40 kg cruising near 25 m/s and achieving a lift-to-drag ratio of roughly 25 requires about 500 W for steady level flight in still air. 

\textbf{Mission payloads:} The power required by mission payloads depends on the operational role of the platform, including communications, sensing, computing, and their integrated use.
\begin{itemize}
    \item \textit{Communication Payloads:} Wide-area access and high-rate backhaul can impose substantial and sustained draws. On medium-class airframes, a multibeam radio-frequency access network, excluding the power amplifiers, typically consumes about 1–2 kW, and the requirement grows further when high equivalent isotropically radiated power is targeted looking ahead, the migration toward terahertz carriers and extra-large MIMO increases spectral efficiency and lowers latency, yet it also raises total energy through denser radio chains, faster data converters, and heavier baseband processing.
    \item \textit{Sensing Payloads:} When sensing is implemented via integrated sensing and communications using the same waveform and RF chains as the communications payload, the two functions largely share the same energy budget. In this regime, incremental sensing cost is minimal and is dominated by signal processing rather than additional radio transmit power. By contrast, carrying dedicated high-precision active sensors such as synthetic aperture radar introduces an extra transmit load on the order of 1–2 kW during illumination intervals. The temporally concentrated nature of these bursts creates pronounced power peaks that must be absorbed by the energy system without degrading station-keeping or ongoing communications.
    \item \textit{Computing Payloads:} Sustained onboard processing that supports edge intelligence, network function virtualization, and multi-sensor data fusion contributes a persistent baseline draw of several hundred watts. As inference and adaptation migrate closer to the platform to satisfy latency and privacy constraints, this term becomes more prominent and less amenable to deep duty-cycle reduction.
\end{itemize}
These elements shape the energy profile of a HAP. As missions expand toward higher bandwidths, larger sensing footprints, and deeper onboard intelligence, aggregate demand will rise. This intensifies the coupling between propulsion provisioning and payload scheduling, and it underscores the value of coordinated energy management across time, across modalities, and across platforms.

\subsection{Storage in Stratospheric Grid}
At night, when solar input is unavailable, a HAP must rely entirely on its onboard energy storage system to sustain lift, propulsion, and mission payloads. Consequently, the energy harvested during daylight must be efficiently stored to support continuous operation over several nighttime hours. As storage, it also accumulates daytime energy via advanced batteries or regenerative fuel cell systems and releases it on demand to buffer transient peaks, support contingencies and eclipse periods, and stabilize power delivery for propulsion and mission loads.

Because the platform is highly mass-sensitive, the storage subsystem must achieve exceptionally high specific energy. Otherwise, even a small increase in battery mass can trigger a chain of structural and aerodynamic penalties. Heavier batteries demand stronger and heavier wings and airframes, which in turn raise the lift and thrust required for steady flight, further increasing discharge power and overall mass. This self-reinforcing cycle has led to the conceptualization of the HAP as a “flying battery,” with a system-level specific energy target typically exceeding 500 Wh/kg. To satisfy this demanding constraint, research has converged on two main technological pathways, including advanced secondary batteries and regenerative fuel cell systems (RFCSs). Advanced secondary batteries, such as lithium–sulfur and solid-state lithium systems benefiting from the high theoretical capacity of sulfur, can reach a theoretical specific energy around 500 to 700 $\text{ Wh/kg}$. On the other hand, RFCSs enable day–night energy transfer through an electrochemical cycle that couples electrolysis, gas storage, and fuel-cell conversion. NASA’s studies on lightweight RFCS architectures for solar-powered high-altitude aircraft report a system-level specific energy of approximately 790 Wh/kg with an electrical efficiency of about 53.4\%, enabling support for payloads of several hundred kilograms and extended night endurance for missions such as Earth observation, resource mapping, and communications relay.

\textbf{In summary}, HAP possesses the intrinsic capacity to operate as energy generation, load, and storage. Guided by mission requirements, it may combine these roles concurrently or only play some of them, and engage in the grid function through WPT to interconnect platforms and orchestrate energy routing, in order to satisfy task-level performance targets. The IGGLS reveals the holistic structure of a stratospheric grid as shown in Fig. \ref{fig:1}. Within this architecture, the stratospheric grid enables dynamic energy flow, real-time balancing of surpluses and deficits across platforms, and coordinated scheduling of power with communications, sensing, and computing tasks. It improves continuity of service through on-demand buffering, supports contingencies and eclipse periods, and enhances resilience by reallocating generation where and when it is most valuable. The result is an energy-cooperative HAP network with higher availability, greater mission flexibility, and strengthened endurance at the constellation scale.

\begin{table*}[htbp]
\centering
\caption{Different Sizes of HAPs Parameters}
\label{tab:haps_power_analysis}
\renewcommand{\arraystretch}{1.1}
\footnotesize
\setlength{\tabcolsep}{2.8pt}
\begin{tabular}{@{}p{2.0cm}p{2.0cm}p{2.0cm}p{1.6cm}p{2.0cm}p{2.0cm}p{1.8cm}p{1.8cm}@{}}
\toprule
\textbf{Category} &
\textbf{PV Area (m²)} &
\textbf{Station-Keeping (kW)} &
\textbf{Comm./Sens. (kW)} &
\textbf{Compute (W)} &
\textbf{Payload (kg)} &
\textbf{Endur. (h)} &
\textbf{Produce (kW)}\\
\midrule
Small (Zephyr-class) &
5--10 &
0.2--0.5 &
0.01--0.05 &
$\le$ 20 &
5--8 &
$>$ 24 &
1.1-2.5\\
\rowcolor[gray]{0.95}
Medium (Aquila-class) &
30--50 &
3--5 &
1--2 &
$\approx$ 100 &
40--50 &
24--36 &
6.6-12.5\\

Large (Sceye-class) &
80--90 &
$\approx$ 14 &
3--5 (RF)+1--2 (SAR) &
200--300 &
200--250 &
$>$ 72 &
17.6-22.5\\
\bottomrule
\end{tabular}
\end{table*}

\section{The Stratospheric Grid in a Broader Ecosystem}
Although the stratospheric grid built upon HAPs enables efficient intra-layer energy utilization, confining energy flows solely within the stratospheric domain still leads to underused daytime surpluses and limited support during nocturnal or high-demand periods. To achieve truly sustainable and balanced energy circulation, the stratospheric grid must be further explored as an active participant in broader multi-layer energy ecosystems. This involves reexamining its role not only as an autonomous energy network but also as an integral component within cross-domain energy exchange, interfacing with space-based, aerial, and terrestrial infrastructures to enhance overall efficiency, resilience, and service continuity.

\subsection{Powering Terrestrial and Aerial Systems}
The stratospheric grid can leverage its abundant solar resource and wide-area visibility to act as a persistent, renewable energy supplier. Rather than letting daytime overproduction dissipate, the stratospheric grid can perform as the energy source to downlink energy to devices whose supply is often intermittent or limited. Typical beneficiaries include internet of things (IoT) sensor networks deployed in remote or oceanic regions, low-altitude drone fleets conducting continuous surveillance or logistics, and emergency or post-disaster communication nodes where terrestrial infrastructure is compromised. By providing directed wireless power, the stratospheric layer closes the gap between energy availability and energy demand across vertical domains, enabling a new form of on-demand aerial energy provisioning.

To achieve efficient supply, adaptive WPT scheduling and spatial energy multiplexing become key. Multiple HAP nodes can coordinate beamforming directions and power levels based on predicted ground demand and environmental conditions, ensuring maximum utilization of surplus generation. At the network level, intelligent control planes driven by artificial intelligence (AI) can perform predictive energy routing, matching energy-rich nodes with target receivers through dynamic priority mapping. Beyond basic power delivery, this architecture enables mission-aware energy orchestration, where energy supply is aligned with task timing and priority. For instance, intensifying energy delivery to UAV fleets during inspection missions or reducing broadcast power during idle phases. The innovation lies in transforming isolated stratospheric grid into active stratospheric energy broadcasters, realizing a distributed, renewable, and demand-responsive energy ecosystem that bridges the gap between sky and ground.

\subsection{Space and Ground Energy Intake}
Although intra-stratospheric energy sharing enables HAP-to-HAP replenishment, the stratospheric grid remains constrained by limited daytime generation and universal reliance on storage at night. In periods of low irradiance, during high-latitude winters, or under peak traffic, aggregate demand can exceed available supply and inter-platform transfers may be neither sufficient nor timely. Sustained operation therefore requires supplementary energy from outside the stratospheric layer.

This motivates a bi-directional energy ecosystem in which the stratosphere functions as an active participant in global energy circulation rather than an isolated self-sustaining tier. Terrestrial WPT hubs equipped with large phased-array antennas can beam microwave power to nearby HAP nodes for immediate top-ups. In parallel, space-based solar power systems located in geostationary or medium Earth orbit can transmit laser or microwave energy to stratospheric receivers, creating a multi-layered power corridor that links space-borne generation to aerial operations. Multiple commercial initiatives are converging on this multi-layer concept. Orbiting Grid’s “Synthetic Laser Exchange Module” and Celestia Energy’s 30\,kW Terra transmitter propose direct orbital or ground-to-air laser power distribution \cite{orbgrid-multilayer,celestia-30kw}. In parallel, Exlumina’s Everlight and ORiS’s satellite-laser architectures plan to deploy orbit-to-drone power beaming \cite{exlumina_orbgrid,ORiS_grid}, further validating the technical continuity from space relays down to stratospheric receivers.

Beyond opportunistic replenishment, intelligent coordination enables predictive energy intake. Control planes should anticipate shortfalls using environmental forecasts, traffic load predictions, and orbital geometries, and then proactively align receiver attitude, pointing, and power budgets. Distributed energy-negotiation protocols can also arbitrate intake priorities among HAP nodes to ensure equitable and efficient allocation during multi-source transfers. In this way, energy inflow becomes a controllable network parameter that co-evolves with service demand and establishes the foundation for a hierarchical, cross-domain energy web spanning space, the stratosphere, and the ground.

\subsection{Cross-Layer Energy Relay}
The stratospheric grid functions as an interconnection fabric that establishes and optimizes energy paths across space, air, and ground nodes. It supplies topology, routing, and control for power flows, coordinating relay forwarding and reconfigurable reflection to deliver assured and adaptive energy connectivity. 

\subsubsection{Relay-Based Energy Forwarding}
From a relay perspective, the stratospheric grid operates as a constellation of aerial repeaters that forward energy between heterogeneous sources and receivers. A HAP can act as an intermediate node in two complementary modes. First, in amplify-and-forward operation, it draws on its own surplus generation to regenerate the incoming beam and transmit a higher effective level toward the target, improving delivery when sources and loads are geographically or geometrically separated. Second, in store-and-forward operation, it exploits onboard storage to buffer received energy and release it later on demand, enabling time-shifted replenishment that aligns supply with mission timing and access windows. Together, real-time regeneration and buffered relay extend coverage, raise end-to-end efficiency, and convert transient surpluses into usable service where and when it is needed. Through dynamic routing algorithms, the grid can determine optimal relay paths based on instantaneous link quality, residual energy, and demand distribution, realizing adaptive energy flow control in the vertical dimension. This flexibility establishes the stratosphere as an intermediate energy exchange layer that stabilizes power transfer under nonuniform generation and consumption conditions.

\subsubsection{RIS-Assisted Energy Routing}
Reconfigurable Intelligent Surfaces (RIS) provide another effective means of enhancing the grid’s networking capability. In this context, RIS modules deployed on HAPs or tethered balloons can reshape electromagnetic propagation to redirect and focus wireless energy toward designated receivers. For microwave and millimeter-wave transmission, metasurface-based passive RIS panels can achieve controllable phase manipulation to reduce path loss and shadowing. Alternatively, active RIS equipped with integrated amplifiers can simultaneously compensate for propagation attenuation and redistribute energy over multiple beams, enabling fine-grained multi-user or multi-platform energy delivery. For optical WPT, adaptive mirror arrays or liquid-crystal spatial light modulators can serve as functional optical RIS, dynamically adjusting reflection phase and direction to maintain beam alignment under platform motion or atmospheric turbulence. This optical adaptation is particularly useful for maintaining stable long-range laser energy corridors between spaceborne power stations, stratospheric grids, and aerial receivers.

By integrating relay and RIS functions, the stratospheric grid transcends the limitations of point-to-point transmission. It evolves into a distributed energy-routing network, where intelligent reflection, selective amplification, and adaptive path coordination jointly maximize energy reach and efficiency across multiple atmospheric and orbital layers. Such a design enables the grid to fulfill its true “network” role to provide reliable, adaptive, and topology-aware power connectivity across space, air, and ground systems.

\subsection{Distributed Aerial Energy Storage}
Beyond its functions as source, load, and relay, the stratospheric grid inherently forms a distributed energy storage layer suspended in the atmosphere. These storage units collectively constitute a cooperative reservoir, capable of stabilizing supply and compensating for spatial or temporal mismatches between generation and consumption.

Unlike ground-based storage farms, the stratospheric layer benefits from low latency, wide coverage, and direct visibility to both space and terrestrial infrastructures. This allows energy packets to be buffered at altitude and selectively dispatched downward or laterally according to forecasted demand. In emergency scenarios such as post-disaster recovery or communication outages, stored energy in the stratospheric grid can be redirected to sustain critical aerial relays or reestablish power to isolated surface networks, effectively functioning as an energy continuity layer between domains.

Moreover, integrating predictive control and federated energy management enables these airborne reservoirs to self-organize into a virtual energy pool. Each node negotiates storage utilization based on residual capacity, expected insolation, and mission priority, achieving decentralized but coordinated operation. In this sense, the stratospheric grid transcends its physical role as a collection of platforms. It evolves into an adaptive, energy-aware infrastructure that not only stores energy but also learns when, where, and how to release it for maximal network utility.

\textbf{In summary}, through repositioning the stratospheric grid within the IGGLS paradigm for broader ecosystem applications, it has demonstrated multiple functions including serving as an aerial energy supply platform, a cross layer energy relay node, and a distributed energy storage layer. Together, these multi-role dynamics establish a foundation for a truly integrated space–air–ground energy ecosystem.

\section{CASE STUDIES}
To validate the advantages of the proposed stratospheric grid framework, a 24-hour cooperative energy operation scenario involving three stratospheric HAPs was constructed. The simulation is based on realistic stratospheric environmental parameters, including typical solar irradiation cycles, atmospheric conditions, task load distributions, and available external energy supplementation. Specifically, the peak solar irradiance be assumed to be  1250 W/m². All HAPs are equipped with 50 m² photovoltaic panels and novel secondary batteries with a specific energy of 700 Wh/kg, alongside WPT transceiving capabilities. The mission payloads and battery capacities differ among the HAPs. HAP 1, designated as a source node, carries lighter task loads and greater storage capacity, enabling significant energy surplus. HAP 2, acting as a load node, carries heavier mission payloads and smaller storage to reduce hovering losses. HAP 3, with moderate task loads and storage capacity, can flexibly switch between source and load roles based on its operational status. To address potential system-wide energy deficits during periods of high-load operation or nighttime functioning of the HAP cluster, a high-power ground-based transmitter was additionally deployed to supply external energy replenishment. Furthermore, to prevent damage from deep discharge and ensure emergency power supply under extreme conditions, a 20\% state-of-charge (SOC) safety threshold is set for all onboard batteries. The objective of the energy dispatch strategy is to minimize the cluster's total schedulable energy, defined as the sum of energy supplied by the ground network to cover the cluster's energy deficit and the surplus energy generated within the HAPs cluster itself.

\begin{figure*}[htbp]
    \centering
    \begin{subfigure}[b]{0.45\textwidth}
        \centering
        \includegraphics[width=\linewidth]{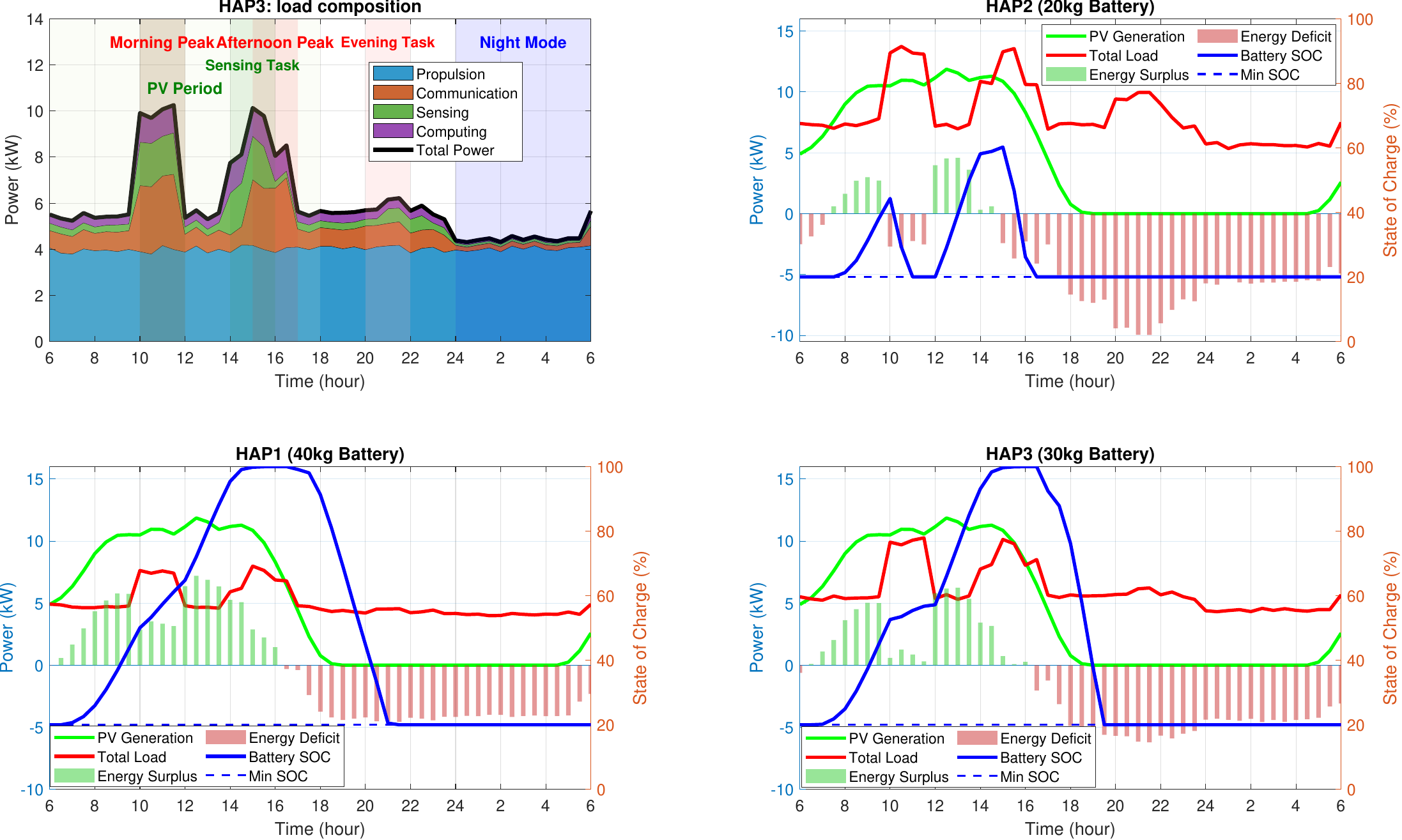}
        \caption{Conventional independent operation}
        \label{fig:sub1}
    \end{subfigure}
    \hfill
    \begin{subfigure}[b]{0.45\textwidth}
        \centering
        \includegraphics[width=\linewidth]{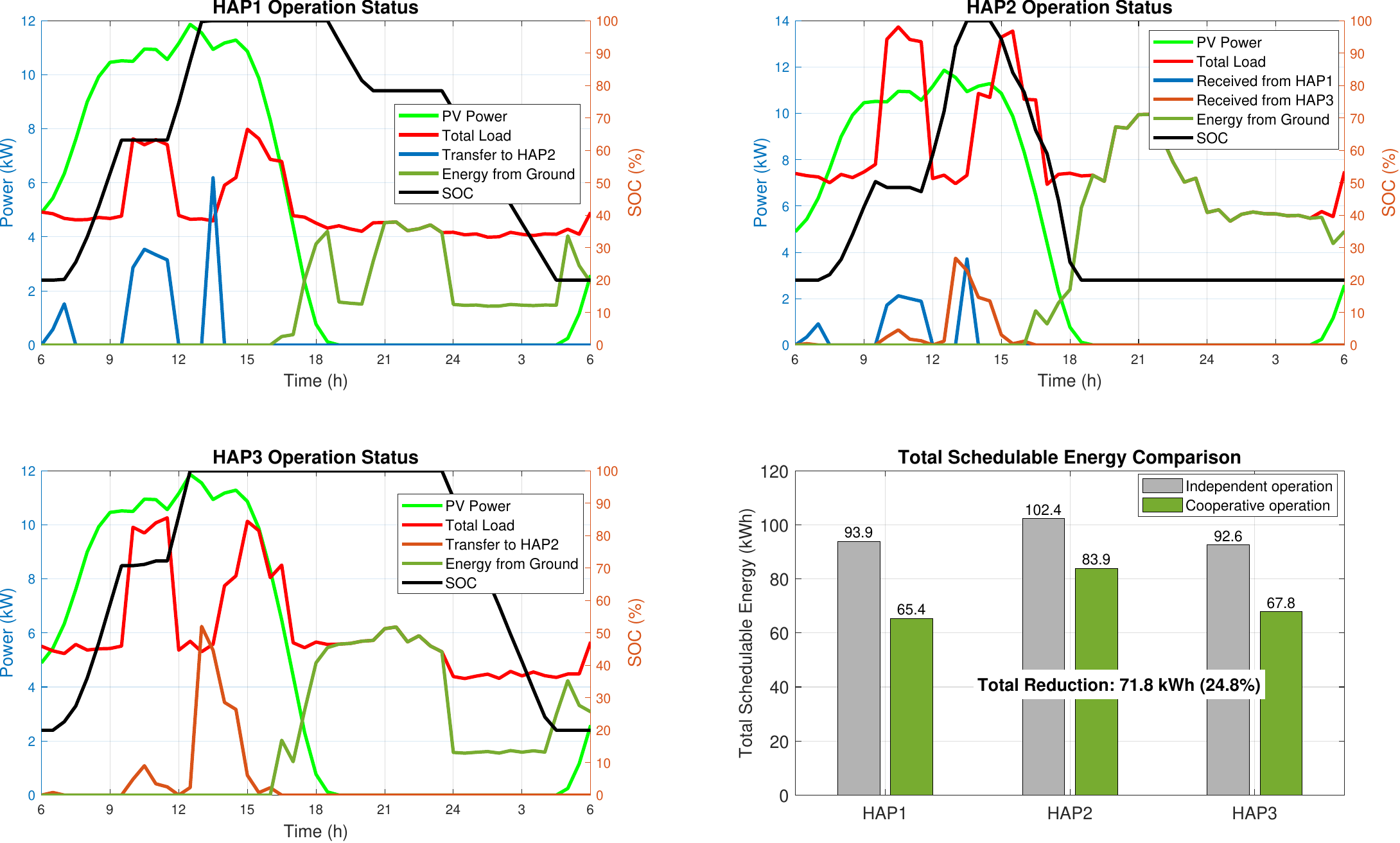}
        \caption{Stratospheric Grid-enabled cooperative operation}
        \label{fig:sub2}
    \end{subfigure}
    \caption{Optimal energy scheduling between three HAPs}
    \label{fig:combined}
\end{figure*}

The effectiveness of the proposed joint operation method for the stratospheric grid is verified by comparing it against a traditional independent operation mode. As shown in Fig.~1(a), the total operational load of the HAPs closely correlates with mission activity. During daytime hours with sufficient solar irradiance, HAP 1 and HAP 3 exhibit significant energy surpluses. In contrast, HAP 2, burdened with heavier mission loads, continues to experience an energy deficit during high-load periods, and its limited storage cannot be effectively charged, leading to low storage utilization. During nighttime, devoid of photovoltaic generation, all three HAPs relying solely on their batteries cannot sustain normal operations until dawn, exhibiting power shortfalls of varying degrees. Regardless of the time of day, the power deficits of the independently operating HAPs cluster necessitate supplementation from the ground network, substantially increasing the cluster's total schedulable energy and highlighting the limitations of the independent operation paradigm.

With the introduction of the stratospheric grid cooperative mechanism in Fig.~1(b), HAP 1 and HAP 3 actively transfer energy to HAP 2 during periods of abundance. This not only compensates for HAP 2's daytime deficit but also charges its battery, enhancing HAP 2's energy storage utilization. In the nighttime phase, all HAPs operate as coordinated storage nodes. By optimizing the charge-discharge strategies while respecting the 20\% minimum SOC safety threshold, the system not only manages nighttime power supply but also provides emergency support when conventional ground-based power is constrained. Simulation results demonstrate that this cooperative mechanism effectively utilizes the surplus energy from HAP 1 and HAP 3 and the storage capacity of HAP 2, reducing the system's total schedulable energy by 24.8\%.

This case study confirms that the proposed energy dispatch mechanism not only ensures the stable operation of heterogeneous HAPs clusters but also significantly reduces the cluster's total schedulable energy through internal energy cooperation. The system demonstrates excellent resilience, particularly when confronting sudden load fluctuations or local energy shortages. When an individual HAP faces an energy deficit, neighboring nodes can provide timely support, effectively preventing service interruptions typical of traditional single-node systems. This “energy sharing" paradigm offers a crucial reference and a scalable technical framework for the energy management of future large-scale stratospheric platforms. The modular design of the system allows for the further integration of additional renewable energy and storage technologies, laying a solid foundation for building fully autonomous stratospheric infrastructure.

\section{Challenges And Future Research Directions}
While the novel architecture and energy dispatch schemes proposed for the stratospheric grid demonstrate significant potential, advancing the stratospheric grid from proof-of-concept to engineering application necessitates overcoming core issues inherent in its key technological pillars, such as intelligent beam PAT, cooperative energy optimization, and system level secure transmission. This section analyzes critical open problems within the current framework and outlines promising research directions aimed at enhancing the system's efficiency, scalability, and resilience.

\subsection{Intelligent Beam PAT in High-Dynamic Environments}
The paramount challenge in achieving stratospheric WPT lies in maintaining stable beam pointing under the high-frequency disturbances caused by platform motion and atmospheric turbulence. Laser beams are susceptible to atmospheric turbulence, leading to beam broadening, wander, and scintillation. In contrast, microwave beams offer better penetration, their narrow-beam characteristics also impose extremely high requirements on pointing accuracy. Existing studies have achieved a certain degree of beam pointing stabilization through high-bandwidth feedback control and sophisticated sensor fusion technologies. However, these methods still exhibit significant limitations when dealing with wide-frequency composite disturbances, as the response delay of feedback systems makes it difficult to fully track high-frequency jitter components, and sensor noise with modeling errors further restricts compensation accuracy under complex atmospheric conditions. As a result, existing systems often suffer from unstable dynamic links and sharp declines in transmission efficiency in practical applications.

Future research should focus on developing intelligent adaptive PAT systems. This involves utilizing data from fused multi-sensor systems (e.g., Inertial Measurement Unit, optical beacons) and deep learning models to predict and feed-forward compensate for platform jitter and turbulence effects in real-time. Concurrently, exploring the deep integration of adaptive optics with intelligent control is a critical path to break through the current technological bottleneck, enabling precise energy transmission that is nearly static despite the dynamic environment. These techniques can enable real-time stabilization even with dynamic platform motion and intermittent atmospheric distortion~\cite{nwpu_dronenews, darpa_wavefront}.

\subsection{AI-Driven Cooperative Energy Optimization}
The stratospheric grid is essentially a spatially distributed energy internet, whose efficient operation hinges on solving cooperative optimization problems involving multiple agents and timescales. The system must achieve optimal global energy scheduling under strong uncertainties from stochastic renewable energy, dynamic loads, and complex network topology. Traditional optimization methods face dual constraints of model mismatch and computational complexity when addressing this high-dimensional complex system.

AI technologies, represented by large-scale pre-trained models and multi-agent reinforcement learning, offer a potential paradigm shift to address this challenge. Future research should focus on constructing foundational models for the power grid. Leveraging massive data from digital twins, these models would possess deep cognitive capabilities to understand the system's physical constraints and operational rules. Building upon this, developing distributed autonomous decision-making frameworks enables individual nodes to make coordinated decisions based on local information and the global situation, achieving optimal energy dispatch. This new paradigm, combining data-driven and model-guided approaches, is key to realizing system-level efficiency and resilient operation.

\subsection{Electromagnetic Compatibility and System Security}
As an open critical infrastructure, high-power WPT links, while carrying energy, also represent potential sources of strong electromagnetic interference and physical layer security vulnerabilities. Beyond these technological risks, the concentrated energy beam also raises critical biosafety concerns for airborne organisms. Consequently, the energy beam could be maliciously jammed or stolen, its electromagnetic emissions might compromise the normal operation of sensitive electronic equipment, and unintended exposure could pose hazards to biological life.

Building a system-level security and resilience defense requires coordinated efforts across the information, physical, and electromagnetic domains. Research should aim to design integrated air-ground cooperative control protocols featuring authentication and encryption, and explore using blockchain technology to establish trusted cross-domain transaction mechanisms. On the electromagnetic compatibility front, it is essential to clarify the interference mechanisms between high-power transmission and sensitive equipment, and develop strategies for active suppression and intelligent spectrum management. Promoting the establishment of relevant international standards and safety operating procedures is the fundamental prerequisite for ensuring the system's secure interoperability and scalable application.



\section{Conclusions}
This paper introduces a stratospheric power grid, a novel architecture operating under the IGGLS paradigm that establishes an interconnected energy network within the stratosphere through WPT. In this system, each HAP node can flexibly switch between power source and load roles based on real-time energy status and mission requirements, enabling autonomous energy dispatch and optimal allocation within the stratosphere.
In addition, this article explores the potential application of stratospheric power grids in a wider range of ecosystems, emphasizing the advantages of stratospheric grids in improving the overall operational capability of the ecosystems. This enhanced capability paves the way for supporting demanding future services, particularly in the 6G communication era. Simulation analysis shows that the stratospheric grid architecture effectively optimizes the energy supply and demand between HAP nodes. This greatly improves the continuous operation capability and energy utilization efficiency of the system.

\bibliographystyle{IEEEtran}
\bibliography{reference,manualrefsEros,references}

\end{document}